\documentclass[sn-mathphys-num]{sn-jnl}

\usepackage{graphicx}%
\usepackage{multirow}%
\usepackage{amsmath,amssymb,amsfonts}%
\usepackage{amsthm}%
\usepackage{mathrsfs}%
\usepackage[title]{appendix}%
\usepackage{xcolor}%
\usepackage{textcomp}%
\usepackage{manyfoot}%
\usepackage{booktabs}%
\usepackage{algorithm}%
\usepackage{algorithmicx}%
\usepackage{algpseudocode}%
\usepackage{listings}%
\usepackage{verbatim}
\newcommand{\rev}[1]{{\color{black} #1}} 

\theoremstyle{thmstyleone}%
\newtheorem{theorem}{Theorem}
\newtheorem{proposition}[theorem]{Proposition}%

\theoremstyle{thmstyletwo}%
\newtheorem{example}{Example}%
\newtheorem{remark}{Remark}%

\theoremstyle{thmstylethree}%
\newtheorem{definition}{Definition}%

\begin{document}

\title[Article Title]{An Advanced Hybrid Quantum Tabu Search Approach to Vehicle Routing Problems}


\author*[1,2]{\fnm{James B.} \sur{Holliday}}\email{jbhollid@uark.edu}

\author[3]{\fnm{Eneko} \sur{Osaba}}\email{eneko.osaba@tecnalia.com}

\author*[1]{\fnm{Khoa} \sur{Luu}}\email{khoaluu@uark.edu}

\affil*[1]{\orgdiv{EECS Department}, \orgname{University of Arkansas}, \orgaddress{ \city{Fayetteville}, \state{AR}, \country{USA}}}

\affil[2]{\orgname{J. B. Hunt Inc.}, \orgaddress{\city{Lowell}, \state{AR}, \country{USA}}}

\affil[3]{\orgname{TECNALIA, Basque Research and Technology Alliance (BRTA)}, \orgaddress{48160 Derio, \country{Spain}}}


\abstract{Quantum computing (QC) is expected to solve incredibly difficult problems, including finding optimal solutions to combinatorial optimization problems. However, to date, QC alone has still not been able to demonstrate this capability except for small-sized problems. Hybrid approaches, where QC and classical computing work together, have shown the most potential for solving real-world-scale problems. This work aims to show that we can enhance a classical optimization algorithm with QC so that it can overcome this limitation. We present a new Hybrid Quantum-classical Tabu Search (HQTS) algorithm to solve the Capacitated Vehicle Routing Problem (CVRP). Based on our prior work, HQTS leverages QC for routing within a classical tabu search framework. The quantum component formulates the Traveling Salesman Problem (TSP) for each route as a Quadratic Unconstrained Binary Optimization (QUBO), solved using D-Wave’s Advantage system. Experiments investigate the impact of quantum routing frequency and starting solution methods. While different starting solution methods, including quantum-based and classical heuristics methods, it shows minimal overall impact. HQTS achieved optimal or near-optimal solutions for several CVRP problems, outperforming other hybrid CVRP algorithms and significantly reducing the optimality gap compared to preliminary research. The experimental results demonstrate that more frequent quantum routing improves solution quality and runtime. The findings highlight the potential of integrating QC within meta-heuristic frameworks for complex optimization in vehicle routing problems.}




\maketitle

\section{Introduction}\label{sec1}

The current era of Quantum Computing (QC) has brought significant attention. Even outside the immediately related fields, many industries are interested in QC. Industries where quantum could create significant disruption if the quantum advantage is achievable, and the transportation industry is one of them. Many transportation industry problems are considered some of the hardest known problems to solve, including NP-hard problems such as the Traveling Salesman Problem (TSP) and the Vehicle Routing Problem (VRP). QC boasts that it will be able to solve problems like these much better and faster than classical approaches \cite{abbas2023quantum}.

Although TSPs and VRPs have been extensively studied for many years, they remain an interesting problem due to their direct connection to the transportation industry. Transportation companies like J.B. Hunt Inc. must address these issues daily as part of their standard operations, and QC research is already focusing on these problems \cite{osaba2022systematic}. In simple terms, the TSP aims to find a tour or route for a salesperson to visit a distinct set of locations while traveling the shortest distance possible. The main difference between the TSP and the VRP is that while the TSP creates a single route to visit all the locations, the VRP allows multiple routes to be used to visit all the locations. These different routes are represented as distinct vehicles. There are many ways to find reasonable solutions to these problems, and achieving near-optimality is still valuable for many industry use cases. However, in this work, we aim to discover whether we can find optimal solutions using a hybrid QC algorithm.

This era of QC is most commonly known as the Noisy Intermediate-Scale Quantum (NISQ) era. It is called noisy because the quantum computers of this era produce high error rates. Intermediate-Scale means these same computers do not yet contain many qubits. Most research around solving industry problems relies on toy versions, so the limited number of qubits is sufficient to map the complete problem onto quantum computers. \rev{A full CVRP formulation would demand \(O(n^2)\) variables, exceeding 10,000 qubits for a problem with just one hundred locations.} Other research points to a different direction, where hybrid quantum/classical algorithms are used so that the problems can scale larger. Hybrid algorithms allow portions of the problem to be solved classically, while other portions can be solved using the quantum computer. It can be done so that only some of a given problem needs to be mapped at any given time, reducing the number of qubits required for the quantum computation. 

\textbf{Our Contributions in this Work:} 
In this article, we explore ways to enhance a hybrid quantum/classical algorithm by strengthening its quantum components. The proposed algorithm will be further extended from our preliminary research work \cite{holliday2024tabu}, where we introduced a hybrid algorithm to solve the CVRP. The main refinements made over the previous version of the method include the alternative starting solutions and the frequency of quantum routing. Our refined algorithm significantly improves the preliminary results and outperforms other hybrid quantum algorithms, achieving a lower overall deviation from optimal solutions. Notably, the optimal and near-optimal solutions are found for several problems, with one case being only 0.03\% away from the optimum. To achieve these results, we investigate the impact of varying the frequency of quantum routing within a tabu search (TS) algorithm. New results show that more frequent use of QC leads to faster convergence to better solutions. We also examine the effect of different starting solutions, including classical heuristics, clustering techniques, and quantum-generated solutions. While results suggest that simpler classical heuristics can be effective starting points, our study does not conclusively demonstrate superiority over quantum-generated solutions. We provide runtime data for the algorithm on various problem sizes, offering insights into the computational cost of the hybrid approach. We highlight the limitations of other hybrid approaches for the VRP, particularly when dealing with realistic datasets where optimal routes do not conform to simple cluster structures. Our proposed algorithm demonstrates better performance in these scenarios.

The remainder of this article is organized as follows. Section \ref{sec2} will review the background related to this research, including formal definitions. Section \ref{sec3} will introduce our algorithm and explain how it was enhanced based on our preliminary research. Section \ref{sec4} will focus on our experimental setup and results. Finally, Section \ref{sec5} will contain the conclusions.

\section{Background}\label{sec2}
In this section, we provide definitions, including the capacitated vehicle routing problem and adiabatic quantum computing. We then discuss ways to solve optimization problems classically and using QC. Lastly, we explore hybrid approaches to solving the CVRP that use both classical and QC methods.

\subsection{Definitions}\label{subsec2}
\subsubsection{Vehicle Routing Problem (VRP)}\label{subsubsec2}
The VRP is formally defined as an NP-hard combinatorial optimization problem with a directed graph \(G = (V, E)\), where \rev{\(V = \{0,1,...,n\}\)} is the set of locations and \(E = \{ (v_i,v_j): v_i,v_j\in V, i \ne j)\}\) is the set of edges between the locations. Location \(v_0\) is not a delivery location but is the depot where the routes begin and end. The set or fleet of vehicles is \(K\). We will discuss constraints around vehicles in the next section. There is also a cost matrix over the edges \rev{\(C = (c_{ij})_{n \times n}\)}. The cost is defined as the distance between two locations in the graph. 

In order to solve the VRP, we must create at most \(|K|\) routes that start and end at the depot location and minimize total cost. For simplicity, we will say the set \(N\) is equal to \(V\), except that it does not contain the depot, \(v_0\). \(x_{ijk}\) is defined as the binary variable \(x\) for an edge between location \(i\) and location \(j\) on vehicle \(k\). \rev{Let \(u_{ik}\) be an integer variable representing the visit position of location \(i\) on vehicle \(k's\) route.} We can mathematically represent this problem as a minimization problem with the constraints as follows,
\begin{align}
 & \rev{min \sum_{k\in K} \sum_{i\in V} \sum_{j\in V} c_{ij} x_{ijk}}
 \end{align}
 s. t. 
 \begin{align}
 & \sum_{k\in K} \sum_{j\in V} x_{ijk} = 1 \quad \forall i \in V\\
 & \sum_{j\in N} x_{0jk} = 1 \quad \forall k \in K\\
 & \sum_{i\in N} x_{i0k} = 1 \quad \forall k \in K\\
 & \sum_{i\in N} x_{ijk} - \sum_{i\in N} x_{jik} = 0 \quad \forall j \in N, k \in K\\
 & 2 \leq u_{ik} \le N \quad \forall i \in N, k \in K\\ 
 & u_{ik} - u_{jk} + 1 \leq (N - 1)(1 - x_{ijk}) \quad \forall i, j \in N, k \in K\\
 & x_{ijk} \in \{0,1\}
\end{align}
In this equation, (1) is the minimization function subject to additional constraints. Constraint (2) ensures that each location is visited by only one vehicle. Constraints (3) and (4) ensure that each route on each vehicle starts and ends at the depot. Constraint (5) ensures that the number of entries into a location equals the number of exits.  \rev{Constraints (6) and (7) prevent sub-tours using the Miller-Tucker-Zemlin formulation \cite{miller1960integer}, ensuring routes are connected to the depot without disconnected cycles.} Constraint (8) defines the primary binary decision variable. 

\subsubsection{Variants of the Vehicle Routing Problem}\label{subsubsec2}

\rev{The CVRP is the VRP variant we focus on in this research. The CVRP introduces a constraint, such that} each vehicle's assigned route must adhere to a capacity constraint \(Q\). Therefore, the sum of each location on the route's demand \(q\) must be less than or equal to the vehicle's value for \(Q\). Homogeneous fleet and heterogeneous fleet variations exist for a fixed value of \(Q\) or a dynamic value of \(Q\) per vehicle. In this research, \(Q\) is a fixed value for each vehicle. 
\begin{equation} \label{eqn:xjik}
 \sum_{i\in V} q_i \sum_{j\in V} x_{ijk} \leq Q \quad \forall k \in K\\
\end{equation}
The constraint in Eqn. \eqref{eqn:xjik} limits the customers on each vehicle \(k\) so that the capacity of the vehicle is not exceeded. This constraint is added to the list of constraints from the previous section to fully define the CVRP.   

\subsubsection{Adiabatic Quantum Computing} \label{subsubsec2} 

Today, we can distinguish between two types of real quantum devices, including gate-based quantum systems and quantum annealers. On the one hand, a gate-based computer uses qubits to perform simple quantum circuit operations, similar to the classical operations on regular bits. It may be joined in any order to form algorithms. A common term for this form is \textit{universal quantum computer}.

On the other hand, a quantum annealer operates on the principle of adiabatic computation, where an initially simple Hamiltonian is gradually evolved from its ground state to the ground state of a final, problem-specific Hamiltonian. If the Hamiltonian evolves slowly enough, the adiabatic theorem ensures that the system stays in the ground state throughout the entire computation. A quantum annealing process starts by representing the problem as an energy landscape. At the outset, the quantum system is set up in a superposition of all potential solutions, corresponding to a high-energy state. As the system progresses, the Hamiltonian is modified, enabling the system to navigate the energy landscape. The objective is to steer the system towards the lowest energy state, representing the optimal solution to the problem. 

In quantum annealers, the adiabatic theorem is deliberately relaxed, permitting the system to evolve faster than the adiabatic limit would typically allow. Consequently, transitions to high-energy states often happen along the evolution. To address this situation, innovative methods for reaching adiabaticity have been introduced in the literature \cite{Takahashi_2017, ferreirovélez2024shortcutsadiabaticvariationalalgorithms}. 

Finally, although this computational model is also universal \cite{PhysRevLett.99.070502}, the D-Wave quantum annealer, employed in this research work, relies on an Ising Hamiltonian, which limits the types of problems that can be executed on the device. Nevertheless, this type of device is particularly suited for solving combinatorial optimization problems \cite{yang2023survey}.

\subsection{Classical Optimization}\label{subsec2}

Route planning is a prominent topic in artificial intelligence due to its significant scientific and social implications. First, these problems are of great scientific interest because they often involve high computational complexity. As NP-Hard problems, solving them presents a significant challenge for the scientific community. Secondly, routing problems are typically designed to address real-world scenarios in logistics and transportation, making their efficient resolution beneficial both socially and commercially.

Numerous methods have been proposed in the literature to tackle these problems efficiently. The most successful methods include exact, heuristic, and meta-heuristic methods \cite{salhi2022overview}. Among them, arguably, meta-heuristics are the most popular methods \cite{hussain2019metaheuristic}. \rev{Meta-heuristics explore the solution space to obtain effective optimization results without depending on the specifics of the problem. It makes them especially well-suited for addressing real-world problems with complex formulations, as they do not require detailed problem-specific information to explore the feasible solution space.

Furthermore, meta-heuristics can also be categorized into search-based algorithms and constructive algorithms. Search-based techniques begin with an initial complete solution or a set of complete solutions, then are modified until a final solution is reached. In contrast, constructive algorithms begin with a partial solution or a set of partial solutions, incrementally built until a complete solution is achieved.

Literature features a wide variety of meta-heuristics. As the primary aim of this research is not to provide an in-depth description of these kinds of algorithms, we recommend readers refer to \cite{ecbestiary} for a comprehensive list of over 100 meta-heuristic approaches.}

\subsection{Quantum Optimization}\label{subsec2}

QC signifies a groundbreaking advancement in computation, leveraging principles from quantum physics to handle information in entirely new ways. Currently, the field is highly anticipated because of its potential to address problems that classical computers find insurmountable, particularly in cryptography, drug discovery, and optimization. This paper concentrates explicitly on the optimization aspect.

Today, several approaches are being investigated in quantum optimization, with quantum annealing (QA) \cite{morita2008mathematical} and variational quantum algorithms, such as the quantum approximate optimization algorithm (QAOA) \cite{farhi2014quantum}, being among the most notable. Despite significant progress, quantum computers are still in their early stages compared to classical computers. They face challenges in efficiently solving problems due to their capacity and inherent instability. As a result, we are in the \textit{noisy intermediate-scale quantum} (NISQ) era \cite{preskill2018quantum}, characterized by the limitations of these devices in effectively handling complex problems.

Despite these challenges, recent studies have increased in number over the past few years, focusing on solving real-world problems using QC. This increase in publications highlights the community's growing interest in exploring the applications of quantum devices. Several factors have contributed to this intriguing development. Two of these factors have likely had the most influence, including the advancements in QC democratization \cite{seskir2023democratization} and the development of increasingly larger and better-connected devices. A primary example of the latter is D-Wave Systems' \texttt{Advantage\_System}, which comprises 5,616 qubits arranged in a Pegasus topology \cite{boothby2020next}. This system is currently the most widely used for solving optimization problems.

Turning our attention to the topic addressed in this paper, research focused on routing problems using QC has been highly prolific, with academic problems such as the Vehicle Routing Problem (VRP) \cite{golden2008vehicle} and the Traveling Salesman Problem (TSP) \cite{matai2010traveling} being the most extensively studied cases. Notably, the survey by Osaba et al. \cite{osaba2022systematic} highlights that this new paradigm has inspired 53 research publications as of 2022. According to the authors of that study, ``\textit{it is noticeable that the TSP engages most of the researchers (60,37\% - 32 out of 53 papers), while the VRP amounts to 25,52\% of the contributions (13 out of 53). The rest of the papers deal with other routing problems, such as the Shortest Path Problem or the Hamiltonian Cycle}''. The trend continues similarly, with VRP and TSP remaining the predominant focus of scientific research in the years following 2022. These problems account for the majority of publications, as evidenced by studies such as \cite{le2023quantum,spyridis2023variational,qian2023comparative,leonidas2023qubit,mohanty2023analysis,sinno2023performance,tambunan2023quantum}.

Analyzing the current body of work reveals two conclusions. First, many studies aim to uncover the potential of quantum technologies or evaluate the efficiency of specific methods, often using academic problems, such as the TSP or VRP, for benchmarking. Second, numerous studies explore the application of QC to real-world routing problems. The latter research category seeks to maximize the capabilities of current NISQ-era devices by implementing efficient and advanced hybrid resolution methods. Representative examples of this trend can be found in \cite{osaba2024solving,weinberg2023supply}.

\subsection{Quantum/Classical Hybrid Algorithms for the VRP}\label{subsec2}

As of today, the most recognized article in this field is \cite{feld2019hybrid}. They focused on solving the CVRP with a two-phase approach. They experimented with quantum in both phases but found the best results when only using it in the second phase. They presented results for problems from the same dataset we experimented with, and we show the difference between their results and ours in Section \ref{sec4}. 

As \cite{feld2019hybrid} is an inspirational work, we will describe their algorithm in detail. For the first assignment phase, they implemented a classical clustering algorithm where each cluster represents a vehicle route. They demonstrated two methods for selecting the starting location for a cluster core, either a location with very high demand or one far from the depot. A cluster is built out from the starting location by adding the closest location to the geometric center of the cluster and then recalculating the geometric center after the location is added. They continue to add the nearest locations to the geometric center until the vehicle capacity is exceeded if another location is added. They then repeat this process with a new starting location until all locations are in a cluster. They then perform cluster improvement where they move locations between clusters if the distance from a location to a cluster center can be decreased by moving it to a different cluster and the capacity of the vehicle \rev{is not} exceeded. \rev{This concludes the first phase of the algorithm.}

The second phase, or routing phase, of the heuristic involves formulating the clusters as a QUBO, allowing the TSP to be solved for each one. The TSP QUBO is formulated as defined in \cite{lucas2014ising}. We present the mathematical formulation for the QUBO because we utilize this same QUBO in our algorithm. Here we alter the notation from \cite{lucas2014ising} and continue with \rev{\(i,j\) defined as before as the edge from location \(i\) to location \(j\), \rev{and \(u\) is the position in the visit sequence on the route. Thus \(x_{i,u}\) is the binary variable indicating location \(i\) is visited at position \(u\).} \(N\) contains all the locations being routed and \(n\) is the number of locations in \(N\). \(E\) is defined as before as the set of edges between the locations.
\begin{equation}\label{eqn:A}
 H_A = A \sum_{j = 1}^N \left( 1 - \sum_{u = 1}^n x_{j,u}\right)^2 + A \sum_{u = 1}^N \left(1 - \sum_{j = 1}^n x_{j, u}\right)^2 + \\ A \sum_{(i,j) \notin E} \sum_{u = 1}^n x_{i,u}x_{j,u+1}
\end{equation}
Here \(H_A\) in Eqn. \eqref{eqn:A} is the QUBO formulation for the Hamiltonian Cycle Problem. The first term ensures that every location appears in the cycle. The second term ensures a $u$th node in the cycle for each $u$. The third term ensures that an edge must exist from $i$ to $j$. The summation over \((i,j) \notin E\) denotes pairs (i, j) not in E.
\begin{equation} \label{eqn:B}
 H_B = B \sum_{(i,j) \in E} c_{ij} \sum_{u = 1}^{n}x_{i,u}x_{j,u+1}
\end{equation}
\(H_B\) in Eqn. \eqref{eqn:B} ensures the cost of the Hamiltonian Cycle is minimized. \(c_{ij}\) again is the cost to travel from location \(i\) to location \(j\).}
\begin{equation} \label{eqn:C}
 H = H_A + H_B
\end{equation}
Adding \(H_A\) and \(H_B\) together in Eqn. \eqref{eqn:C} provides the complete Hamiltonian \(H\), which is the QUBO that will solve the TSP, or what is commonly referred to as the routing phase of the problem. The penalty coefficients are set with \(A\) being higher than the most significant cost in \(C\) and \(B\) being set to 1. 

Reference \cite{borowski2020new} presents a more recent approach to solving the CVRP, introducing multiple hybrid algorithms. The two most performative algorithms were the DBScan Solver (DBSS) and the Solution Partitioning Solver (SPS). DBSS operates very much like \cite{feld2019hybrid}. SPS can use DBSS as a starting solution, where it creates a single TSP solution for all the locations, and then SPS divides the TSP into routes to solve the CVRP. They provided their source code, allowing us to recreate their results in our preliminary research. We also experimented with DBSS and SPS to generate the starting solution for our algorithm. It is explained in Section \ref{sec3}.  

Even more recent QC approaches to solving the CVRP include \cite{palackal2023quantum} and \cite{xie2024feasibility}, where hybrid quantum-classical algorithms, QAOA, and variational quantum eigensolver (VQE) are used to solve the problem. \cite{palackal2023quantum} struggled with QAOA, as it was unable to find feasible solutions to toy CVRP problems; however, using VQE was successful. \cite{xie2024feasibility} added constraint-preserving mixers to their QAOA and found feasible solutions, but only on toy problems. \cite{suen2022enhancing} implemented a full CVRP QUBO formulation that they solved on the quantum-inspired Fujitsu Digital Annealer \cite{nakayama2021description}. They demonstrated results on CVRPs with up to 50 locations and achieved an optimality gap of 3.89\% on a 34-location problem. \cite{arino2023adiabatic} presented the most promising recent result, employing a hybrid approach that combines two phases: quantum clustering and classical routing. They also presented results on some of the problems from the same dataset used in our experiments. The result we highlighted was for problem CMT 5, where they outperformed all other hybrid algorithms with an optimality gap of 4.12\%. Otherwise, their results were inferior to ours on all other problems. However, their experiments show the potential of quantum clustering in a two-phase approach, essentially the opposite of what the highly cited \cite{feld2019hybrid} used in their approach.

\section{Our Proposed Method}\label{sec3}
Since NISQ-era quantum optimization is only possible for small-sized problems, we decided to pursue a hybrid approach. We analyzed the experimental results of other hybrid algorithms and noticed none achieved optimal results for the CVRP. Those hybrid algorithms all utilized a two-phase approach to solving the problem. This approach is discussed in \cite{laporte2000classicalheuristics}. We determined that we should explore ways to improve the existing two-phase approaches. We decided a meta-heuristic hybrid approach might be better suited since our goal is to find optimal solutions.

Meta-heuristics have a mechanism to avoid getting stuck in a local optimum. In our approach, we selected TS \cite{glover1998tabu} as the meta-heuristic because it enabled us to effectively incorporate some of the concepts that other hybrid approaches have utilized. Primarily, our approach solves the TSP for each route by utilizing QC. The primary trade-off associated with using a meta-heuristic, in our case, is the algorithm's runtime. The hybrid algorithms we tried and the one heuristic we tried ran much faster, but always found sub-optimal solutions.

\rev{The hybrid design is critical under NISQ limitations to scale our solution to real-world-sized problems. By solving only individual TSP subproblems ($\leq25$ nodes), our approach requires $\leq625$ logical qubits per QUBO \cite{lucas2014ising}, easily embedded on D-Wave Advantage. Annealing time remains fixed at $20~\mu$s per call, independent of global scale.}
\begin{figure}[hbtp]
    \centering
    \fbox{\includegraphics[width=0.8\linewidth]{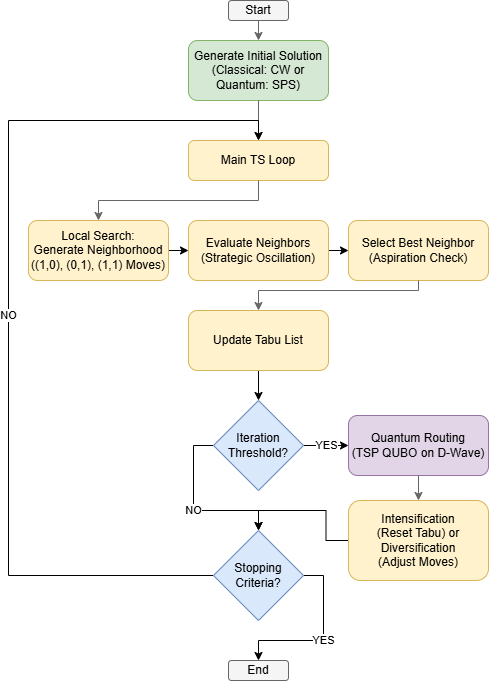}}
    \caption{The proposed flow of HQTS}
    \label{fig:tsflow}
\end{figure}
\subsection{Hybrid Quantum Tabu Search (HQTS)}\label{subsec3}

The research described in \cite{cordeau2002guide} states that TS has historically performed well on the CVRP, reinforcing our decision to create a hybrid quantum/classical TS algorithm. TS has been widely studied for solving optimization problems, and different ways exist to implement it \cite{taillard1993parallel} \cite{gendreau1994tabu} \cite{osman1993metastrategy}. We can break our algorithm into classical and quantum parts at the highest level. The classical part is our implementation of a TS algorithm. The quantum part is a QUBO formulation of the TSP that is solved for each route in a given solution on D-Wave's quantum annealer. We originally named this algorithm Hybrid Quantum Tabu Search. Fig. \ref{fig:tsflow} shows a high-level flowchart for HQTS.

\subsection{Classical Components}\label{subsec3}

TS is a local search optimization technique, with the primary objective of the classical portion of the algorithm to discover the most appropriate assignment of locations to routes. Based on a starting solution, the TS runs in a main loop and performs a local search, making minor changes to that solution and creating a neighborhood of solutions that are all only slightly different from the starting solution. Once the neighborhood is generated, each neighbor solution is evaluated, and the neighbor that leads to the best improvement in the objective function cost is chosen. The chosen solution replaces the starting solution, and the TS returns to the beginning of the main loop, starting the local search again and creating another new neighborhood. It continues until a stopping criterion is met and the best solution found during the search is returned. The name "tabu" is derived from the fact that once a solution is chosen, it is added to a recency memory called the tabu list. While that solution is tabu, meaning it is in the tabu list, it cannot be selected as the next starting solution, except in a specific case we will describe below. 

The tabu list is used to keep track of all recently selected solutions. For our case, the tabu list is considered short-term memory because only recently selected solutions need to stay on the list, and in our implementation, they only stay on the list for some small number of main loop iterations. For our algorithm, we allowed a solution to remain in the tabu list for a random number of iterations, ranging from \(0.4N\) to \(0.6N\), where \(N\) is the number of locations in the problem. This idea of a random length of time came from \cite{taillard1993parallel}. The tabu list serves to prevent the TS from behaving like a greedy search algorithm. It now forces the TS to make less optimal choices at different points in the search, thus driving the search into new parts of the global solution landscape. 

The local search is the most time-consuming part of the algorithm because it must generate solutions and then pick one of them. The neighborhood of solutions is constructed by performing either a (1,0), (0,1), or (1,1) change to the current starting solution. For our method, we define the (0,1) change as swapping a location with another location on the same route. We also define the (1,0) change as moving a location from one route to another. However, the location sequence in the new route is determined by evaluating which spot in the sequence leads to the lowest cost route, as was used in \cite{taillard1993parallel}. A (1, 1) change is swapping two locations from two different routes. Once all possible solutions have been generated by performing all possible changes, each is evaluated by calculating the total cost of each neighbor solution. As we described earlier, the selected solution from the neighborhood is used to create a new starting solution, and the local search is restarted.

There is a condition that can occur, allowing a chosen solution to be selected even if it already exists in the tabu list. This concept is called aspiration \cite{glover1998tabu}. We define aspiration as the point at which the local search discovers a solution that provides the best value yet discovered for the objective function. It then makes the solution the next starting solution, even if \rev{it is} in the tabu list. The search \rev{does not} discard a potentially global optimum solution.

There are other classical components of TS that we implemented. Intensification and diversification are important components \cite{glover1998tabu}. Intensification is the idea of allowing TS to revisit solutions that can search around them more thoroughly. Diversification enables TS to explore previously unseen solutions in the hope of discovering new areas to search. When \(X\) number of main loop iterations happen without finding a new global best solution, we trigger intensification or diversification. \(X\) was selected uniformly between 0.6 and 1.1 times the number of locations in the problem. These values were determined by experimentation. A diversification trigger changes the number of locations to consider when creating new solutions. An intensification trigger resets the tabu list to an empty list. We also extended our intensification with a quantum component, defined in the next section.

Another interesting TS concept we implemented is strategic oscillation (OS) \cite{glover2011case}. It is added because \cite{glover2011case} showed that when TS is forced to search only inside feasible solutions, it can limit its effectiveness. Without OS, the algorithm chooses the best solution that maintains feasibility, but with OS, if the current starting solution is feasible, the algorithm can choose solutions that lead to infeasibility. Once in infeasible territory, the algorithm prioritizes solutions that reduce infeasibility (or worsen it the least) until a feasible solution is found again. This process allows the exploration of a wider solution space.  Alg. \ref{alg:cap} shows how the next starting solution is selected.

\begin{algorithm}
\caption{Solution evaluation with strategic oscillation}\label{alg:cap}
\begin{algorithmic}[1]
\Require {$cbs$ can only be assigned a feasible solution so that it can be evaluated as a global best solution}
\State \rev{$Infeasibilty(N_i)$ returns the total capacity violation for a given candidate solution}
\State $p \gets \textrm{previously selected solution}$
\State $N \gets \textrm{candidate solutions from local search}$
\State $n \gets \textrm{number of candidate solutions}$
\State $cbs \gets unassigned$   \Comment{current best solution}
\State $sbfs \gets unassigned$  \Comment{selected best feasible solution}
\State $sbis \gets unassigned$  \Comment{selected best infeasible solution}
\State $ss \gets unassigned$    \Comment{selected solution}
\If{$IsFeasible(p)$}
    \For{$i \gets 1 \textrm{ to } n$}
        \If{$Cost(N_i) < Cost(sbfs) \textrm{ and } IsFeasible(N_i)$}
            \If{$Cost(N_i) < Cost(cbs)$}
                \State $cbs \gets N_i$      
            \EndIf
            \If{$!IsTabu(N_i)$}
                \State $sbfs \gets N_i$
            \EndIf
        \ElsIf{$Cost(N_i) < Cost(sbis) \textrm{ and } !IsFeasible(N_i)$}
            \If{$\rev{!IsTabu(N_i)}$}
                \State $sbis \gets N_i$
            \EndIf
        \EndIf
    \EndFor
    \If{$Cost(sbis) < \rev{Cost(sbfs)}$}
        \State $ss \gets sbis$
    \Else
        \State $ss \gets sbfs$
    \EndIf 
\Else
    \For{$i \gets 1 \textrm{ to } n$}
        \If{$Infeasibilty(N_i) < Infeasibilty(sbfs) \textrm{ and } IsFeasible(N_i)$}
            \If{$Cost(N_i) < Cost(cbs)$}
                \State $cbs \gets N_i$
            \EndIf
            \If{$!IsTabu(N_i)$}
                \State $sbfs \gets N_i$
            \EndIf
        \ElsIf{$Infeasibilty(N_i) < Infeasibilty(sbis) \textrm{ and } !IsFeasible(N_i)$}
            \If{$!IsTabu(N_i)$}
                \State $sbis \gets N_i$
            \EndIf
        \EndIf
    \EndFor
    \If{$Infeasibilty(sbis) < Infeasibilty(sbfs)$}
        \State $ss \gets sbis$
    \Else
        \State $ss \gets sbfs$
    \EndIf 
\EndIf

\end{algorithmic}
\end{algorithm}

Lastly, we have the stopping criteria. In our preliminary study, we had a hard limit of one hour of wall clock time, or if the main loop \rev{had not} found a new global best solution in the last 5000 main loop iterations. In the extension of this work, we determined to revisit the stopping criteria and allow for longer searches. It has been accomplished by changing the stopping criteria from stopping after 5000 main loop iterations without an improvement to the global best solution to now run for \(N * 100\) main loop iterations without improvement where N is again the number of locations in the problem. We have also removed the one-hour time limit for the wall clock. Section \ref{sec4} contains details on the run time for our algorithm.  

\subsection{Quantum Components}\label{subsec3}
The quantum components have two different objectives, i.e., one to initialize the TS with a starting solution to build from and the other to set the \rev{best} sequence of locations on the routes.  When a TS starts, a solution must be provided to perform a local search. In our experiments, we explored various methods for generating the initial starting solution, including utilizing other quantum hybrid algorithms to create the starting solution. In our preliminary research, we created a naive clustering approach to generate the starting solution. We have developed a more advanced system for generating initial solutions. The most famous heuristic for solving the CVRP is the Clarke-Wright savings heuristic (CW) \cite{clarke1964scheduling}. In our preliminary research, we compared our results to CW and determined that CW produces an average optimality gap of 9.97\% on the dataset we tested. That was better than some of the hybrid algorithms we tested. We tested hybrid algorithms from \cite{borowski2020new} in our preliminary research, and the best of those was only able to produce an average optimality gap of 18.68\%. Still, with the motivation to incorporate more quantum components into our algorithm, we determined that we should test two of the hybrid algorithms and the CW heuristic as methods for generating starting solutions. Section \ref{sec4} contains the results of our experimentation with these different methods.

The other quantum component in our algorithm was in the intensification process. Following \cite{taillard1993parallel}, we considered calculating the routes (solving the TSP for one route) from the best-known solution at certain points during the search. To do it, we save the best solution found throughout the search. When the quantum routing is triggered, we build a QUBO formulation as described in Eqn. \eqref{eqn:A}, Eqn. \eqref{eqn:B} and Eqn. \eqref{eqn:C} for the TSP of each route in the best-known solution and then send the QUBO to the quantum computer for optimization. It is done one by one for each route in the solution. The processing time for the quantum computer to solve these TSP QUBO's is in the order of milliseconds. After the quantum computer is finished, we reformat the results in our classical framework. The classical formulation becomes the next starting solution for the next iteration of the main loop. In \cite{taillard1993parallel}, they would perform the routing step every twenty main loop iterations. 

\rev{This quantum component is not employed to outperform classical TSP solvers in isolation, but to generate high-quality route segments. This allows us to still utilize QA to solve harder CVRPs at a scale unattainable by quantum means alone.} Currently, access to QC is not free and involves cost, so the fewer calls to QC, the lower the cost for our algorithm. Additionally, accessing QC is done using a cloud service, so more calls can increase the probability of an access error. Our preliminary algorithm would trigger this process if TS had not found a new global best solution in the last 2000 iterations of the main loop. In our new approach, we attempted to trigger this form of intensification at varying intervals. Section \ref{sec4} will show how changing that interval impacted our results.

\section{Experiments and Results}\label{sec4}
\subsection{Datasets and Setup}\label{subsec4}
\begin{figure}[t]
\fbox{\includegraphics[width=\linewidth]{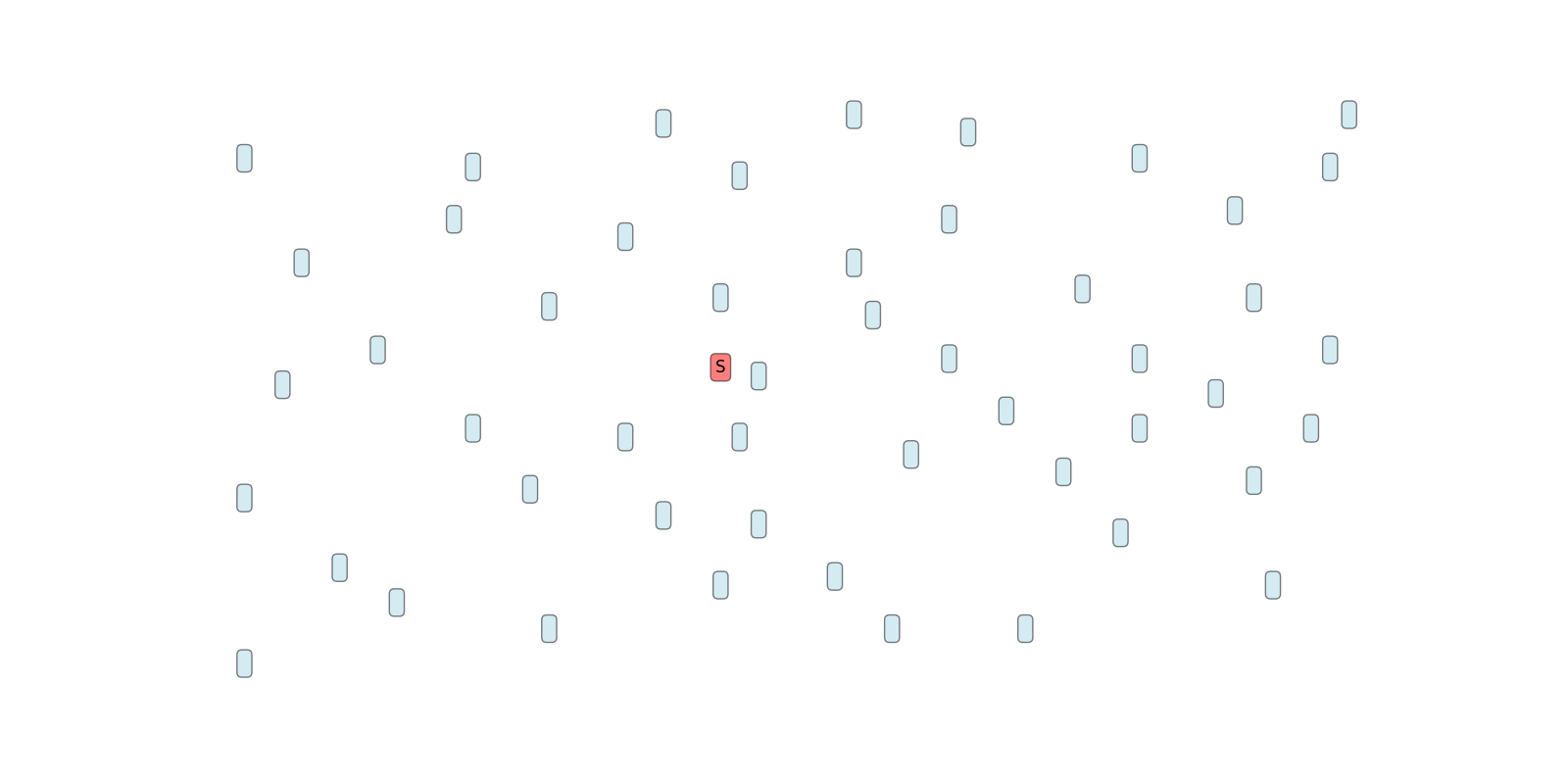}}
\caption{Visualization of CMT 1 (Red = Depot, Gray = Delivery Location)}
\label{fig:cmt1visualization}
\end{figure}

\begin{figure}[t]
    \fbox{\includegraphics[width=\linewidth]{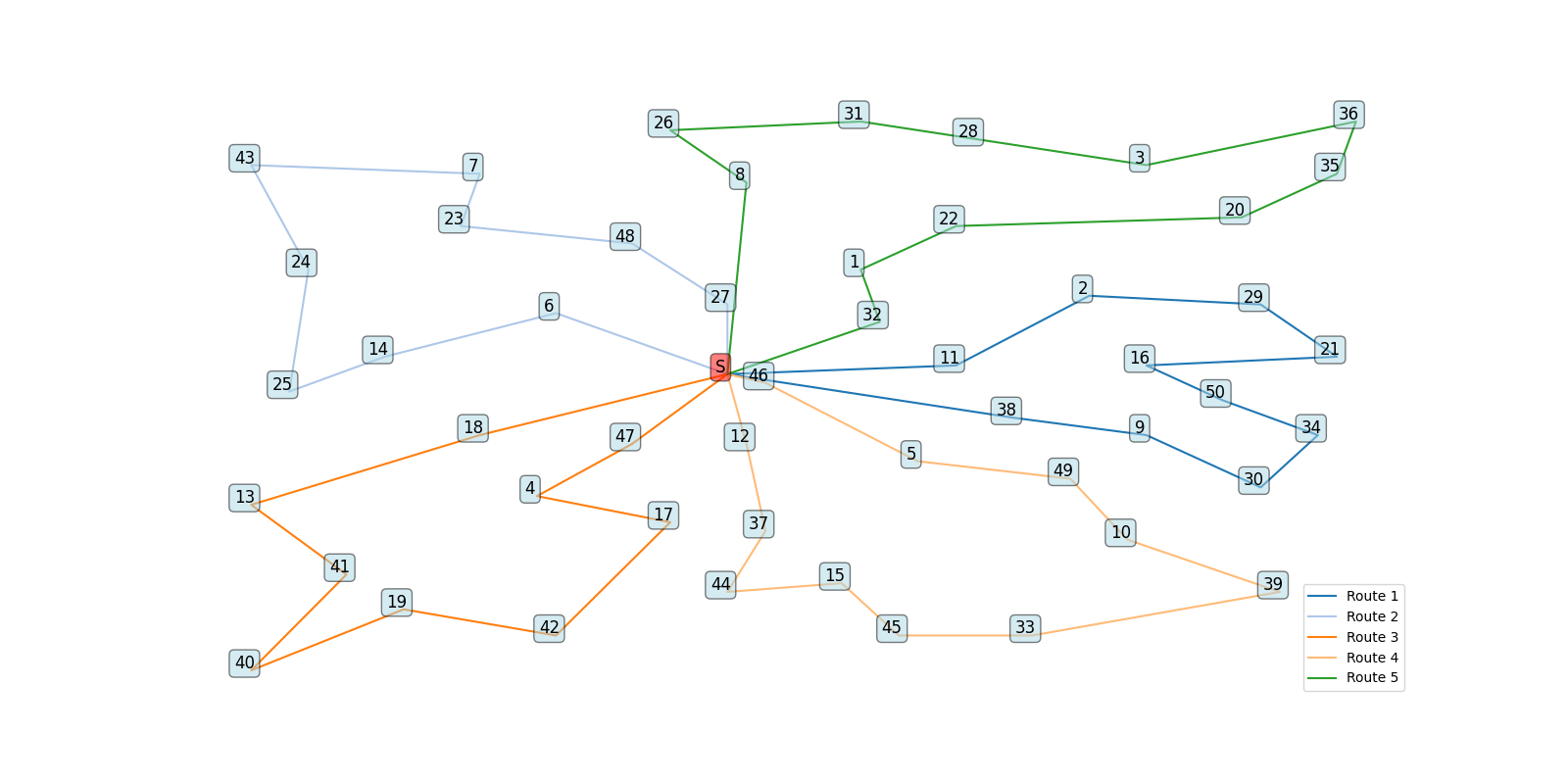}}
    \caption{CMT 1 Optimal Result using HQTS}
    \label{fig:cmt1hqts}
\end{figure}

The extensive research focused on the VRP and CVRP has yielded significant resources, including solutions and benchmarks for evaluating solutions. For our study, we sought a dataset that would provide a solution the industry could find interesting. A dataset with problems that are representative of real-world routing problems. Typically, when routing vehicles around a metropolitan area, you will likely find a depot centrally located among the points it serves. In this scenario, the routes often resemble flower petals in shape, as a route would leave the depot, making deliveries as it makes an arc away and back to its starting location. In some cases, deliveries will be very clustered, but that \rev{is not} the rule. With that in mind, there exists a dataset that serves very well for this type of routing evaluation, and it is the dataset from Christofides, Mingozzi, and Toth (CMT) \cite{cmtdatset}. When visualizing the optimal routes for this dataset, you can often see that flower shape. A visual representation of Problem 1 and a visualization of the optimal solution found from HQTS can be seen in Figs. \ref{fig:cmt1visualization} and \ref{fig:cmt1hqts}. Optimal route visualizations for all of the CMT datasets are available at http://vrp.galgos.inf.puc-rio.br/index.php/en/. 

The CMT dataset was chosen for additional reasons. This kind of dataset provides a substantial benchmark to compare against because of its history, so you can immediately see how your algorithm ranks with other research. We discovered other hybrid algorithms \cite{feld2019hybrid} \cite{borowski2020new} \cite{sales2023adiabatic} that have shown results with this dataset. Another important aspect of this dataset is that the optimal solution to each problem has been found and proven. The dataset consists of 14 problems, with problems 1-5 having varying numbers of locations, ranging from 50 to 199. Table \ref{dataset} presents the details of each problem. Additionally, the locations are randomly located around a central depot. Problems 11 and 12 have more clustered locations, in 11 the depot is centrally located, but in 12 the depot is shifted to one side. Problems 6-10 have the same layout as problems 1-5 but contain an additional time window detail for each location. This applies to problems 13 and 14 as well. We only solved problems 1-5, 11, and 12 because they were explicitly designed as CVRP problems. The other problems are similar but designed for the CVRPTW, which we did not attempt to solve with our algorithm. Although the locations appear to be randomly placed, they effectively represent a real-world problem.

\begin{table*}[h]
    \caption{CMT Dataset}
    \label{dataset}
    \centering
    \scalebox{0.80}{
        \begin{tabular}{lll}
        \hline
        {Problem No.} & {No. of Locations} & {Location Distribution} \\
        \hline
         1 & 50 & Random\\
         2 & 75 & Random\\
         3 & 100 & Random\\
         4 & 150 & Random\\
         5 & 199 & Random\\
         6 & 50 & Random\\
         7 & 75 & Random\\
         8 & 100 & Random\\
         9 & 150 & Random\\
         10 & 199 & Random\\
         11 & 120 & Cluster\\
         12 & 100 & Cluster\\
         13 & 120 & Cluster\\
         14 & 100 & Cluster\\
         \hline
    \end{tabular}
    }
\end{table*}

 All of our experiments were run on a GitHub Codespace \cite{githubcodespaces}. The Codespace had a four-core processor with 16 GB of RAM that was used for the classical computation of our algorithm. The quantum part of the algorithm was performed on D-Wave's Advantage System 4.1 device. We accessed this system via D-Wave's cloud API \cite{dwaveleap}. The \texttt{num\_reads} parameter was set to 1000 for all calls to the QC.  
 
\subsection{Preliminary Results Examined}\label{subsec4}
In our preliminary work \cite{holliday2024tabu}, we compared our solution against other hybrid solutions. Specifically, we compared our results against those reported by the hybrid algorithm in \cite{feld2019hybrid} and the DBSS and SPS algorithms in \cite{borowski2020new}, which we recreated using their provided code. For this problem, CMT 1, our algorithm found the BKS, which, as stated above, is also the optimal solution to the problem. That is the only problem where we achieved optimality. Still, our algorithm outperformed the other hybrid solutions on four of the seven problems we tested in the CMT dataset, resulting in an overall lower percentage deviation from the optimal solution than the other hybrid algorithms. We also compared our results to some classical, well-known heuristics and outperformed them in three of six problems. The heuristics were Clarke-Wright \cite{clarke1964scheduling}, Fisher-Jaikumar \cite{fisher1981generalized}, and Sweep \cite{gillett1974heuristic}. The heuristic's results were reported in \cite{feld2019hybrid}. While \rev{one result was encouraging} when we found an optimal solution, \rev{performance fell short of expectations} in that our algorithm was unable to outperform the other hybrid algorithms on each problem.  

\subsection{\color{blue}Ablation Study on HQTS Configurations}\label{subsec4}
\color{blue}We compare the refined HQTS against a baseline configuration from \cite{holliday2024tabu}, treating prior results as an earlier parameter setting. In this study, \color{black} we compared our new results with our preliminary results and the results reported from \cite{feld2019hybrid} hybrid algorithm because it was the second highest performing hybrid algorithm from our preliminary work. This comparison is illustrated in Table \ref{results1} and Fig. \ref{fig:percent_deviation}. For reference, we also present a comparison of our new results with those of the same classical heuristics we reported in our preliminary research. This comparison is illustrated in Table \ref{results2}. 

For this new research, three main questions drove our experimentation. Could we use more QC in our process, and would doing so improve our results? Could we use QC to generate an initial solution, and how would that compare to using a heuristic approach or clustered approaches for the starting solution? Thereafter, answering the previous two questions, would that guide us to finding better solutions than before? 

\subsubsection{Question 1: Quantum Routing Delay}\label{sec431}
To answer the first question, we set up a test in which we varied the frequency of running the quantum routing part of the algorithm. In our preliminary research, we chose a delay of 2000 iterations of the search without any improvement before performing the quantum routing. The number 2000 was chosen arbitrarily and mostly related to limiting our usage of the quantum computer because we had very limited QC time. We decided to try different values for the delay. The experiments ranged from 3000 to 250 (3000, 2500, 2000, 1500, 1000, 500, 250). To experiment, we ran the algorithm on CMT 1. If the result was equal to the BKS or within a 2\% deviation from optimal, we considered it a good run. Specifically for CMT 1, the allowed solution range was 524.61 to 532.8. We cumulated ten good runs for each of the different values for the delay. We cataloged the average number of iterations to the best solution found and the average wall clock time (run time) in seconds for the ten good runs at each delay value in Table \ref{delays}. 

\begin{figure}[t]
    \centering
    \fbox{\includegraphics[width=0.97\linewidth]{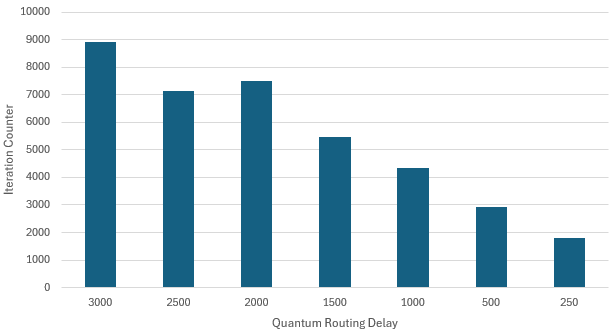}}
    \caption{Average No. of Iterations to Best Solution Found with Different Quantum Routing Delays on CMT1}
    \label{fig:iterations}
\end{figure}

\begin{figure}[t]
    \centering
    \fbox{\includegraphics[width=0.97\linewidth]{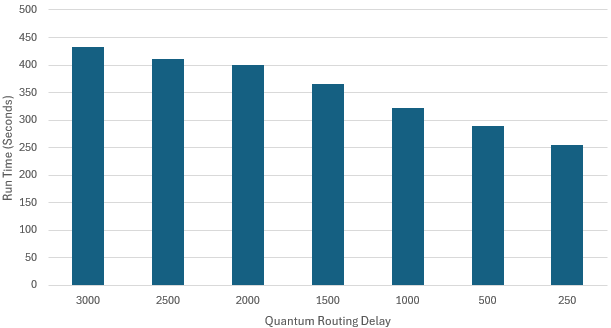}}
    \caption{Average Run Time to Best Solution Found with Different Quantum Routing Delays on CMT1}
    \label{fig:runtime}
\end{figure}

The results show that the best results for both metrics were achieved with a quantum routing delay of 250. This means allowing the quantum routing to trigger more frequently and more quickly when non-improving iterations occur, causing the algorithm to find a better solution faster in terms of both the number of iterations and wall-clock time. Fig. \ref{fig:iterations} shows, on average, the decreasing number of iterations needed to find the best solution, and Fig. \ref{fig:runtime} shows, on average, the decreasing run time in seconds to find the best solution.  

\begin{table*}[h]
    \caption{Average Results For Different Quantum Routing Delays On Problem CMT 1}
    \label{delays}
    \centering
    \scalebox{0.80}{
        \begin{tabular}{lll}
        \hline
        {Quantum Routing Delay} & {Iterations To Best Solution Found} & {Run Time (seconds)} \\
        \hline
         3000 & 8899 & 433 \\
         2500 & 7153 & 411 \\
         2000 & 7510 & 400 \\
         1500 & 5458 & 365 \\
         1000 & 4336 & 322 \\
         500 & 2926 & 290 \\
         250 & 1811 & 255 \\
         \hline
    \end{tabular}
    }
\end{table*}

\subsubsection{Question 2: Starting Solution}
For the second question, we initially used different starting solutions, with a quantum routing delay of 2000. The first option for the starting solution was the clustered approach we used in our preliminary research (refer to that work for an explanation of this approach). The next was using the Clarke-Wright savings heuristic (CW) \cite{clarke1964scheduling}. For the remaining two, we used hybrid quantum approaches. These approaches were introduced in \cite{borowski2020new}, and we recreated them in our preliminary research, specifically in the Solution Partitioning Solver with DBSCAN (SPSDBSCAN) and Solution Partitioning Solver without DBSCAN (SPSFULLQ). 

SPSFULLQ was the best-performing hybrid approach we were able to re-create. However, it is limited to only being able to run on CMT 1 because on more significant problems, that algorithm produces a QUBO that is too large to solve on the DWAVE quantum annealer. The hybrid solutions are noisy, producing different results each time they are run, while clustered and CW produce the same solution every time. We again used CMT 1 for these tests with the same requirements of ten good runs per test. A good run is defined the same as in the previous experiment. Table \ref{starting2k} shows the average results for this experiment. The results here were somewhat surprising in that a more minimal starting solution did not lead to finding a good solution faster.

\begin{table*}[h]
    \caption{Average Performance Comparison For Different Starting Solutions on CMT 1 (Quantum Routing Delay 2000)}
    \label{starting2k}
    \centering
    \scalebox{0.80}{
        \begin{tabular}{llll}
        {Algorithm} & {Starting Solution} & {Iterations To Best Solution Found} & Distance \\
        \hline
         SPS (DBSCAN) & 662.09 & \textbf{3120} & 527.83\\
         SPS (FULLQ) & 714.53 & 5857 & \textbf{524.61}\\
         Cluster & 1108.65 & 7885 & 526.44\\
         CW & \textbf{584.41} & 7510 & 526.54\\         
         \hline
    \end{tabular}
    }
\end{table*}

\begin{figure}[t]
    \fbox{\includegraphics[width=0.97\linewidth]{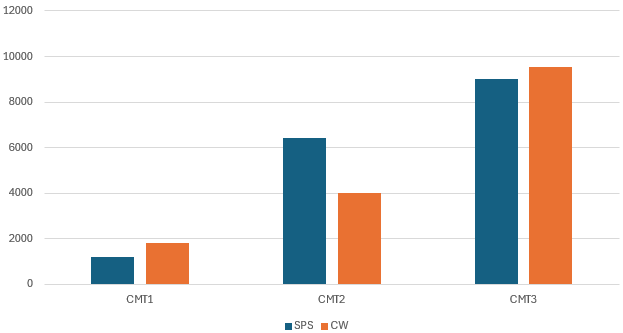}}
    \caption{Average No. of Iterations to Best Solution Found with Different Starting Solutions}
    \label{fig:starting_iterations}
\end{figure}

We then decided to try using the results from the quantum routing experiment in this experiment, so we changed the delay from 2000 to 250, and then repeated the experiment for just SPSDBSCAN and CW. We also extended this experiment into CMT 1, CMT 2, and CMT 3. The results here did not show that either starting solution was consistently better in terms of the time it took to find a better solution. Fig. \ref{fig:starting_iterations} shows that SPSDBSCAN was faster on CMT 1 and CMT 3, but CW was faster on CMT 2. For CMT 2 and CMT 3, we were unable to find the optimal solution. Our best runs used CW, but on average, the results are very close, as shown in Table \ref{starting250}. 

\begin{table*}[h]
    \caption{Performance Comparison For Different Starting Solutions (Quantum Routing Delay 250)}
    \label{starting250}
    \centering
    \scalebox{0.80}{
        \begin{tabular}{llllllll}
        & \multicolumn{3}{c}{SPS (DBSCAN)} & & \multicolumn{3}{c}{CW} \\ 
        \cmidrule{2-4} \cmidrule{6-8} \\
        {Problem} & {Avg. Result} & {Best Result} & {Avg. Iterations} & & {Avg. Result} & {Best Result} & {Avg. Iterations} \\
        \hline
         CMT 1 & \textbf{525.43} & \textbf{524.61} & \textbf{1183} & & 526.54 & \textbf{524.61} & 1811\\
         CMT 2 & \textbf{851.76} & 851.58 & 6401 & & 855.7 & \textbf{848.9} & \textbf{4018} \\
         CMT 3 & \textbf{835.36} & 831.37 & \textbf{9020} & & 836.62 & \textbf{830.99} & 9556 \\
         \hline
    \end{tabular}
    }
\end{table*}

\subsubsection{Question 3: Overall Improvement}
To answer the third question, we applied the answers to questions one and two to the same experiment we conducted in our preliminary research. We ran the tests the same way as before, keeping the best solution from running our algorithm three times. We then adjusted aspects of the algorithm based on the results generated from questions one and two. We used a routing delay of 250 for each experiment and used the CW heuristic to generate the starting solution. We decided to use CW because we found better solutions with the heuristic, even if only slightly, and it would allow us to use less QC in our tests. We wanted to maximize our available QC resources to complete our experiments, and question two's results showed little difference in the starting solution for our algorithm.

\begin{figure}[t]
    \fbox{\includegraphics[width=0.97\linewidth]{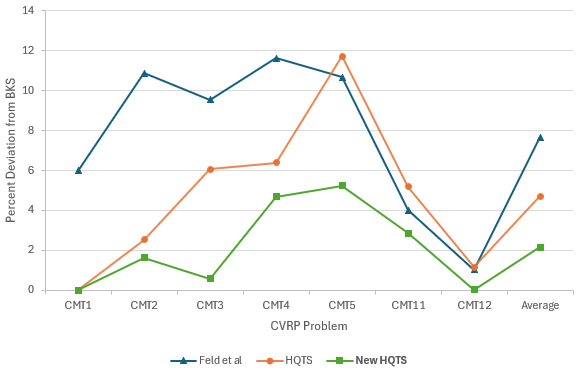}}
    \caption{Percent Deviation from BKS on CMT dataset}
    \label{fig:percent_deviation}
\end{figure}

In Table \ref{results1}, you can see that we matched our results for CMT 1 and improved our results for all the remaining problems. Again, our results show we could only find one optimal solution out of the seven problems. However, for problem CMT 12, our new result was 0.03\% away from the optimal solution. On the five remaining problems, we documented improved results. We found better solutions for each problem, outperforming the hybrid algorithm of \cite{feld2019hybrid}. Overall, we reduced our optimality gap from 4.72\% to 2.15\%. Refer to Fig. \ref{fig:percent_deviation}. Table \ref{results2} is also included to demonstrate how these results compare with those of well-known heuristics on the same problems.

\begin{table*}[h]
    \caption{Performance Comparison of HQTS on the CMT Dataset}
    \label{results1}    
    \scalebox{0.80}{
    \begin{tabular}{llllllllll}
        \multicolumn{2}{c}{}  & \multicolumn{1}{c}{\textbf{  }} & \multicolumn{2}{c}{\textbf{Feld at el Ref. \cite{feld2019hybrid}}} & \multicolumn{2}{c}{\textbf{Preliminary Ref.\cite{holliday2024tabu}}} & \multicolumn{2}{c}{\textbf{New Research}} \\
        \hline
        \textbf{Problem} & \textbf{Size} & \textbf{BKS} & \textbf{Distance} & \textbf{Dev.} & \textbf{Distance} & \textbf{Dev.} & \textbf{Distance} & \textbf{Dev.} \\
        \hline
         CMT 1 & 50 & 524.61 & 556 & 5.98\% & \textbf{524.61} & \textbf{0.0\%} & \textbf{524.61} & \textbf{0.0\%} \\
         CMT 2 & 75 & 835.26 & 926 & 10.86\% & 856 & 2.52\% & \textbf{848.95} & \textbf{1.64\%}\\
         CMT 3 & 100 & 826.14 & 905 & 9.55\% & 876 & 6.06\% & \textbf{830.99} & \textbf{0.59\%}\\
         CMT 4 & 150 & 1,028.42 & 1,148 & 11.63\% & 1,094 & 6.4\% & \textbf{1,076.56} & \textbf{4.68\%}\\
         CMT 5 & 199 & 1,291.29 & 1,429 & 10.66\% & 1,442 & 11.72\% & \textbf{1,359.08} & \textbf{5.25\%}\\
         CMT 11 & 120 & 1,042.12 & 1,084 & 4.02\% & 1,096 & 5.19\% & \textbf{1,071.9} & \textbf{2.86\%}\\
         CMT 12 & 100 & 819.56 & 828 & 1.03\% & 829 & 1.16\%  & \textbf{819.77} & \textbf{0.03\%}\\
         \hline
    \end{tabular}
    }
\end{table*}

\begin{table*}[t]
    \caption{Performance Comparison of well-known heuristics and HQTS on the CMT Dataset}
    \label{results2}
    \centering
        \scalebox{0.7}{
        \begin{tabular}{llllllllllll}
        \multicolumn{2}{c}{} & \multicolumn{1}{c}{} & \multicolumn{1}{c}{} & \multicolumn{2}{c}{\textbf{Clarke-Wright}} & \multicolumn{2}{c}{\textbf{Fisher-Jaikumar}} & \multicolumn{2}{c}{\textbf{Sweep}} & \multicolumn{2}{c}{\textbf{New Research}} \\
        \hline
        & \textbf{Problem} & \textbf{Size} & \textbf{BKS} & \textbf{Distance} & \textbf{Dev.} & \textbf{Distance} & \textbf{Dev.} & \textbf{Distance} & \textbf{Dev.} & \textbf{Distance} & \textbf{Dev.} \\
        \hline
         & CMT 1 & 50 & 524.61 & 585 & 11.5\% & 524 & 0.12\% & 532 & 1.41\% & \textbf{524.61} & \textbf{0.0\%}\\
         & CMT 2 & 75 & 835.26 & 900 & 7.75\% & 857 & 2.6\% & 874 & 4.64\% & \textbf{848.95} & \textbf{1.64\%}\\
         & CMT 3 & 100 & 826.14 & 886 & 7.25\% & 833 & 0.83\% & 851 & 3.01\% & \textbf{830.99} & \textbf{0.59\%}\\
         & CMT 4 & 150 & 1,028.42 & 1,204 & 17.07\% & - & - & 1,079 & 4.92\% & \textbf{1,076.56} & \textbf{4.68\%}\\
         & CMT 5 & 199 & 1,291.29 & 1,540 & 19.26\% & 1,420 & 9.97\% & 1,389 & 7.57\% & \textbf{1,359.08} & \textbf{2.86\%}\\
         & CMT 12 & 100 & 819.56 & 877 & 7.01\% & 848 & 3.47\% & 949 & 15.79\% & \textbf{819.77} & \textbf{0.03\%}\\
    \end{tabular}
    }
\end{table*}

In our preliminary results, we did not report the runtime of the algorithm, so we have decided to include those results as well. Run time \rev{was not} limited at all in these results. Those run times range from 255 seconds on average for the smallest problem of CMT 1 to 37868 seconds on average for the largest problem CMT 5. The average run time for each problem is reported in Table \ref{runtime}.

\begin{table*}[h]
    \caption{Average Run Times on the CMT Dataset}
    \label{runtime}
    \centering
    \scalebox{0.80}{
        \begin{tabular}{lll}
        \hline
        Problem & Size & Run Time (Seconds) \\
        \hline
         CMT 1 & 50 & 255 \\
         CMT 2 & 75 & 1534 \\
         CMT 3 & 100 & 3807 \\
         CMT 4 & 150 & 17687 \\
         CMT 5 & 199 & 37868 \\
         CMT 11 & 120 & 1886 \\
         CMT 12 & 100 & 1169 \\
         \hline
    \end{tabular}
    }
\end{table*}

\section{Conclusions}\label{sec5}
We aimed to show that a hybrid quantum algorithm can find optimal solutions to a real-world optimization problem. While our research identified only one optimal solution and nearly one more, we still demonstrated that QC aids in finding optimal solutions. To date, this is the best-performing hybrid algorithm for the CVRP on the CMT dataset. 

We were able to outperform our previous results in every problem, primarily using QC more often during the experiments. Additionally, using better starting solutions led to improved results. However, we could not demonstrate that a quantum computing-derived starting solution was always superior to a classical heuristic algorithm-derived starting solution. Our average run time was slightly faster, and we found slightly more optimal solutions on average using the quantum starting solutions. However, using the classical heuristic as the starting solution actually found the most optimal solutions for CMT 2 and CMT 3. Finding more ways to use the quantum computer was the main focus of our research, and our new results showed that more QC was usually beneficial.

\rev{Results from Section \ref{sec431} indicate a trend toward improved convergence with reduced routing delay. Although further reductions below 250 iterations may yield additional gains, experimental constraints prevented confirmation of this. Access to QC resources was governed by a fixed quota that could not be adjusted during execution. Exceeding this quota by invoking the quantum solver too frequently would terminate the entire run. Future work will require expanded quantum resource allocation to enable systematic evaluation of delays below 250.}

While the two-phase approach to solving the CVRP has been well studied and the compared hybrid quantum algorithms utilized those ideas \cite{feld2019hybrid} \cite{borowski2020new}, our results showed they \rev{do not} always do well with a dataset like CMT. They show fairly good results on CMT 11 and CMT 12 problems, which is to be expected because the locations are more clustered. They fall behind our results on the randomized CMT 1-5 problems. The shape of the optimal routes was a significant decision point in our algorithm design. \rev{Analysis suggests} the clustering approach for route generation does not do a good job of finding the correct clusters to base the routes on. Because the visualization of the route shape resembles a flower petal, with some stops closer to the depot and others farther away, a clustering algorithm may overlook potentially good locations to add to the route, as they might be closer to the depot and not near other locations in the cluster.

The complexity of the large CVRP problems limited the algorithm's ability to find optimal solutions. \rev{Analysis indicates} the local search part of the algorithm was the main contributor to not finding more optimal solutions. The assignment of locations to routes was more consequential to the suboptimal results than the quantum routing. \rev{The neighborhood generation relied exclusively on basic relocation and swap operators. The operations included only two types of moves: relocating a single customer from one route to another, or swapping two customers across different routes. More sophisticated operators, such as those introduced in \cite{lin1973effective}, were not implemented. Future work will focus on developing a quantum-enhanced local search mechanism to generate higher-quality neighbor solutions with greater efficiency.}

\bibliography{sn-bibliography}

\end{document}